\documentclass[twocolumn,showpacs,preprintnumbers,amsmath,amssymb,superscriptaddress,prb]{revtex4}
\usepackage{graphicx}
\begin{document}
\title{Spin and Charge Currents in SNS Josephson Junction with $\mathbf{f}$-wave Pairing Symmetry}
\author{Yousef Rahnavard}
\address{Department of Physics, Faculty of Sciences, University of Isfahan, 81744 Isfahan, Iran}
\author{Gholamreza Rashedi}
\address{Department of Physics, Faculty of Sciences, University of Isfahan, 81744 Isfahan, Iran}
\author{Takehito Yokoyama}
\address{Department of Physics, Tokyo Institute of Technology, 2-12-1 Ookayama, Meguro-ku, Tokyo 152-8551, Japan}
\date{\today}
\begin{abstract}
We study transport
properties and also local density of states in a clean superconductor-normal metal-superconductor Josephson junction with triplet $f-$wave superconductors based on the Eilenberger equation.
Effects of thickness of normal metal and misorientation between gap vector of superconductors on the spin and charge currents are investigated. Spin current, which stems from the misalignment of d-vectors, in the absence of charge current for some values of phase difference between superconductors is found. We also find unconventional behavior of the spin current associated with 0-$\pi$ transition. 
The misalignment of d-vectors also gives rise to a zero energy peak in the density of states in the normal metal.
\end{abstract}
 \maketitle

\section{Introduction}
\label{SEC1} Superconductivity with spin-triplet pairing symmetry
which has been observed in a series of experiments in
Sr$_{2}$RuO$_{4}$\cite{Maeno,Ishida,Luke,Mackenzie,Nelson,Asano1}, UPt$_{3}$ and some
other heavy fermion
complexes\cite{Tou,Muller,Qian,Abrikosov,Fukuyama,Lebed,Saxena,Pfleiderer,Aoki} attracts
a lot of interests in the condensed matter physics. The spin
triplet $f-$wave pairing symmetry has been considered for
 unconventional superconductivity in UPt$_3$ heavy-fermion
\cite{Graf,Machida,Lussier}. Superconductivity has been observed in
UPt$_3$ compound in two important states, $A-$phase and $B-$phase.
Although different phases correspond to the different
symmetries of the superconducting gap on the Fermi surface,  all
of them belong to spin triplet. The $f-$wave order parameter
has a more complex dependence on azimuthal angles in comparison
with the well-known $p-$wave order parameter for superfluid phases
of $^{3}$He. The Fermi surface of UPt$_3$ is highly complex with
five or six complicated shape sheets. While the pairing mechanism
of UPt$_{3}$ is unknown, a large number of experimental and
theoretical works have been done on its thermodynamic and transport property. The Josephson junction between triplet superconductors is an accurate
method to study physics in triplet superconductors.
In superconductor-normal metal-superconductor (SNS) junction, the Josephson current flows owing to the proximity effect.

Recently, proximity effect in unconventional superconductor and
normal metal junction has attracted much
attention\cite{Tanaka,Tanaka2,Tanuma,Asano2,Yokoyama2,Sawa}. The
proximity effect in junctions between triplet superconductors and
normal metals in the diffusive regime have been investigated
theoretically \cite{Tanaka2,Tanuma,Asano2}. It is found that the
proximity effect is enhanced by the mid-gap Andreev bound state
formed at the interface between triplet superconductor and the
normal metal due to the sign change of the pair potential.
\cite{Tanaka2} As a result, the density of states in the normal
metal has a zero energy peak rather than gap. To investigate the
proximity effect, the quasiclassical Green's function method is
quite useful. In this paper, we apply the quasiclassical
Eilenberger equation\cite{Eilenberger} to calculate charge- and
spin-currents and density of states in a triplet SNS junction.

The SNS Josephson effect is an interesting issue because of
the industrial applications such as superconducting quantum
interference devices (SQUID) \cite{Guttman,Lindell}. DC
Josephson junction has a persistent current by gradient of the macroscopic phase of superconductivity \cite{Golubov}. While the charge current is more known property in
Josephson junction, spin current is more interesting because of applications in spintronics. The effect of phase
difference on the spin current in the spin-triplet SIS Josephson
junction has been studied recently \cite{Kolesnichenko,Asano3}. Authors
of the paper \cite{Kolesnichenko} obtained a polarized
dissipation-less supercurrent of spins. It is shown that the
current-phase dependencies are quite different from those of the junction between conventional
$s$-wave superconductors \cite{Kulik,Rashedi} and high temperature $d$-wave superconductors \cite{Coury}.

In the present paper, we investigate spin and charge Josephson currents, and the density of states in junction between misorientated crystal of triplet $f-$wave superconductors which sandwich the mesoscopic normal metal layer. Triplet $f$-wave pairing symmetry is considered, for example, for
superconductivity in UPt$_3$ crystals \cite{Heffner}.\\

\begin{figure}[tbp]
\includegraphics[width=1.2\columnwidth]{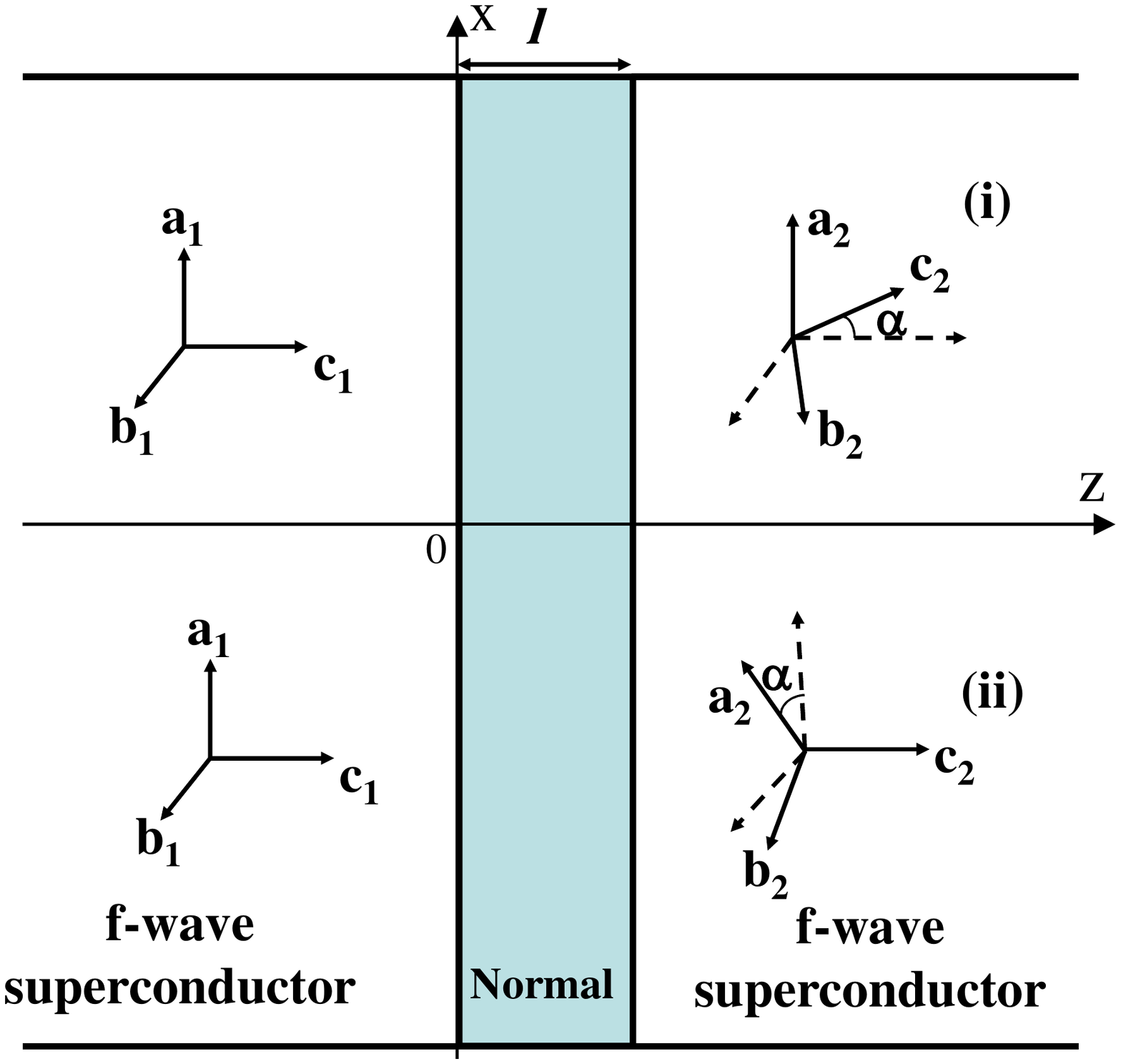}
\caption{(Color online) Scheme of a flat interface between two
superconducting bulks. Here, 1, 2 label the left and right
half-spaces. Two superconducting bulks which are separated by a
layer of normal metal have a misorientation as much as $\alpha$.
Two different geometries are corresponding to the different
orientations of the crystals in the right and left sides of the
interface. In geometry (i) the $bc-$plane in the right side has
been rotated around the $a-$axis by $\alpha$. In geometry (ii),
the $ab-$plane in the right side has been rotated around the
$c-$axis by $\alpha$.  For both of geometries (i) and (ii), we
consider a rotation only in the right superconductor and the
crystallographic $a-$axis in the left half-space is selected
parallel to the partition between the superconductors ($x-$axis).}
\label{fig1}
\end{figure}

The organization of the rest of this paper is as follows. In
Sec.\ref{2} the quasiclassical equations for Green's functions are
presented. Green's functions are obtained in Sec.\ref{3}. The
obtained formulas for the Green's functions are used to calculate
the charge and spin current densities and also density of states in the normal region. In Sec. \ref{4}, numerical results of charge and spin currents in the
normal metal between superconductors are analyzed. Effects of
normal layer thickness between superconductors and misorientation between $d$-vectors of superconductors on the charge and spin currents, and density of
states have been investigated. The paper will be
finished with the conclusions in Sec.\ref{5}.

\section{Formalism and Basic equations}\label{2}
 In this section, we consider a  clean normal metal such as Copper between two misorientated
 $f-$wave superconductors. A normal metal
 layer with the thickness of $l$ is sandwiched by two triplet superconductors. Interfaces of normal metal
 and superconductors have been considered as totally transparent. For the case $l\gg\lambda_{F}$
 we can use the ballistic quasiclassical Eilenberger equation \cite{Eilenberger}
\begin{equation}
\hbar\mathbf{v}_{F}\cdot\nabla \breve{g} +\left[ \varepsilon
_{m}\breve{\sigma}_{3}+i \breve{\Delta},\breve{g}\right] =0,
\label{Eilenberger}
\end{equation}
and the normalization condition $ \breve{g}\breve{g}=\breve{I}$,
where $\varepsilon _{m}=\pi k_BT(2m+1)$ are discrete Matsubara
energies with $m=0,1,2...$, $T$ is the temperature,
$\mathbf{v}_{F}$ is the Fermi velocity and
$\breve{\sigma}_{3}=\hat{\tau}_{3}\otimes \hat{I}$ in which
$\hat{\tau}_3$ is the Pauli matrix in particle-hole space.
$\hat{\sigma}_{j}\left(j=1,2,3\right)$  denote Pauli matrices in
spin space in the following.

\begin{figure}[tbp]
\includegraphics[width=\columnwidth]{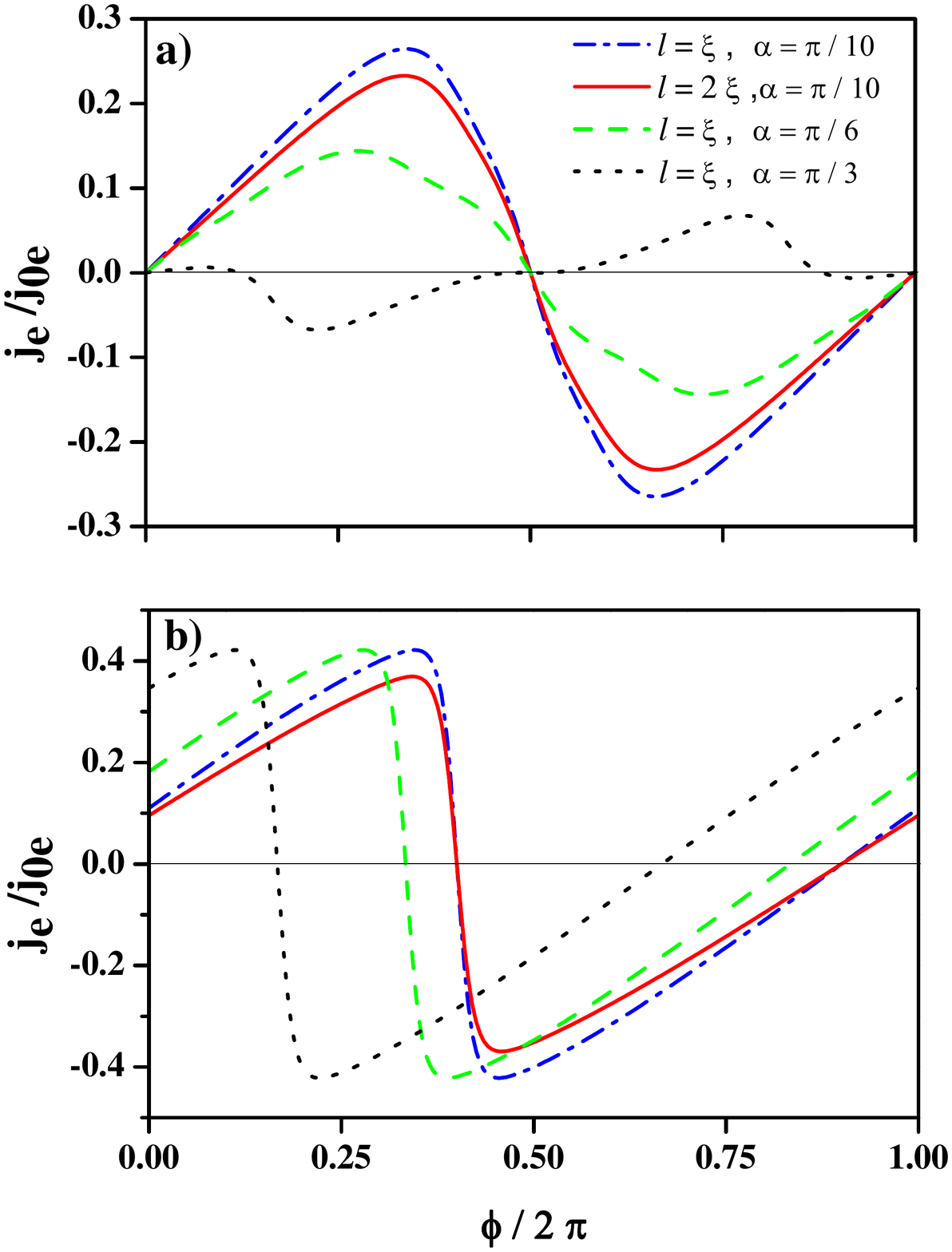}
\caption{(Color online) The normal component ($z$ component) of
the charge current versus the phase difference $ \protect\phi $
for the axial state (\ref{axial}), $T=0.05T_{C}$, different
thicknesses of normal metal and different misorientations between
superconductors. Panel (a) is for geometry (i) and panel (b) for
geometry (ii). Currents are given in units of
$j_{0,e}=\frac{\protect\pi }{8}eN(0)v_{F}\Delta(0) $.}
\label{fig2}
\end{figure}

The Matsubara propagator $\breve{g}$ can be written in the
standard form:
\begin{equation}
\breve{g}=\left(
\begin{array}{cc}
\mathbf{g}_{1}\mathbf{\hat{\sigma}}+g_{1} & \mathbf{g}_{2}\hat{%
\mathbf{\sigma }}i\hat{\sigma}_{2} \\
i\hat{\sigma}_{2}\mathbf{g}_{3}\hat{\mathbf{\sigma }} &
\hat{\sigma}_{2}\mathbf{g}_{1}\hat{\mathbf{\sigma
}}\hat{\sigma}_{2}-g_{1}
\end{array}
\right),
\end{equation} \label{Green's function}
where the matrix structure of the off-diagonal self energy $\breve{%
\Delta}$ in the Nambu space is
\begin{equation}
\breve{\Delta}=\left(
\begin{array}{cc}
0 & \mathbf{d}\hat{\mathbf{\sigma }}i\hat{\sigma}_{2} \\
i\hat{\sigma}_{2}\mathbf{{d^{\ast }}\hat{\sigma}} & 0
\end{array}
\right) .
\end{equation}
\label{order parameter}
In this paper, we focus on the unitary states,
($\mathbf{d\times d}^{\ast }=0$).  Also, we use the
Eilenberger equation for $\mathbf{d}=0$ in the normal
metal region ($0\leq z\leq l$). From the Eilenberger equation it
is clear that Green's functions in the normal metal, $g_{1N}$ and
$\mathbf{g}_{1N}$, are constant for $0\leq z \leq l$. Solutions of
Eq.(\ref{Eilenberger}) has to satisfy the conditions for Green's
functions at the bulk of
superconductors $\breve{g}\left( \pm \infty \right)=\frac{\varepsilon _{m}\breve{\sigma}%
_{3}+i\breve{\Delta}_{2,1}}{\sqrt{\varepsilon _{m}^{2}+\left| \mathbf{d}%
_{2,1}\right| ^{2}}}$.

\begin{figure}[tbp]
\includegraphics[width=\columnwidth]{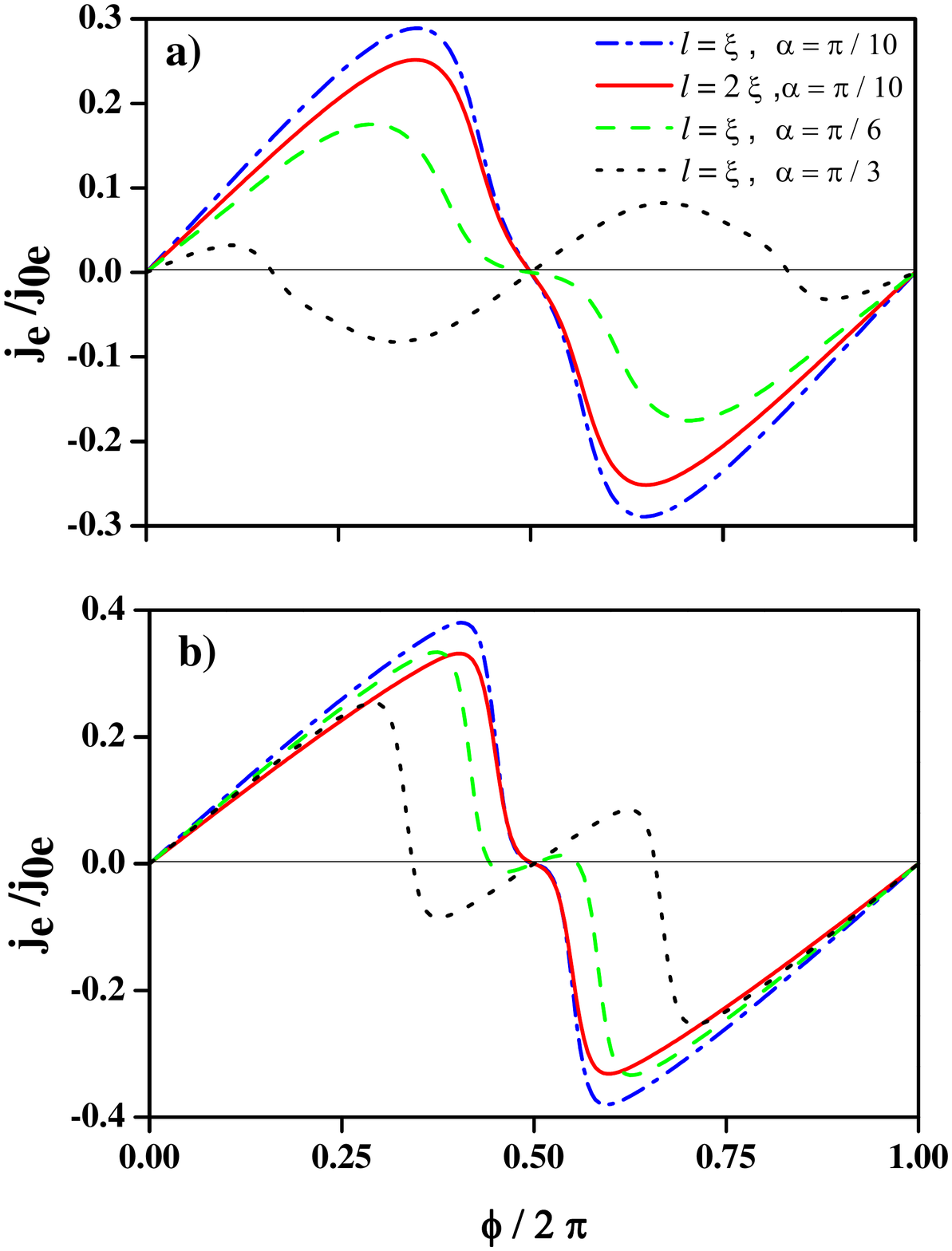}
\caption{(Color online) The normal component ($z$ component) of
the charge current versus the phase difference $ \protect\phi $
for the planar state (\ref{planar}), $T=0.05T_{C}$, different
thicknesses of layer normal metal between superconductors and
different misorientations between superconductors. Panel (a) is
for geometry (i) and panel (b) for geometry (ii). } \label{fig3}
\end{figure}
\begin{figure}[tbp]
\includegraphics[width=1.1\columnwidth]{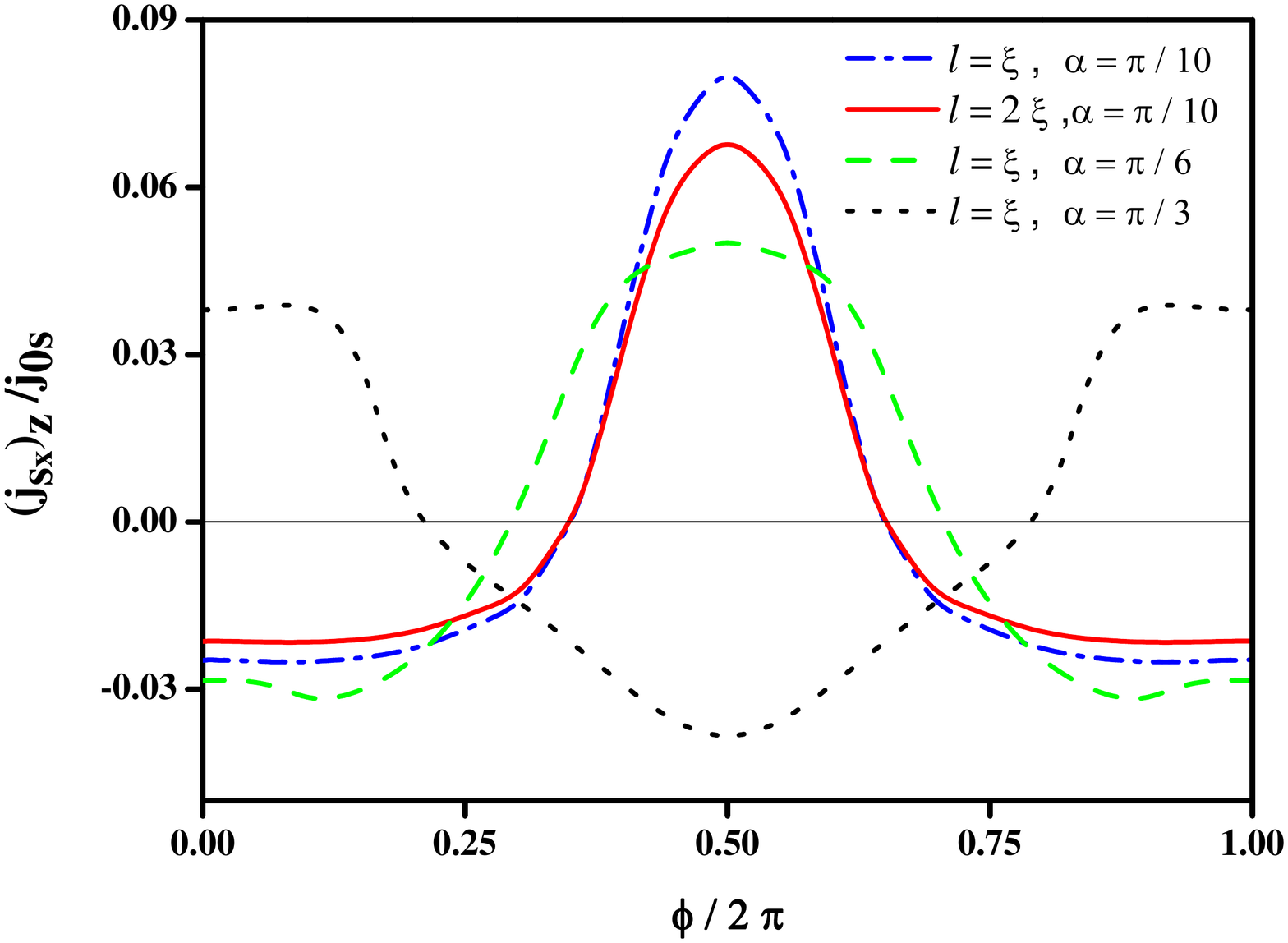}
\caption{(Color online) The normal component of the spin current
$(j_{sx})_z$ versus the phase difference $ \protect\phi $ for the
axial state (\ref{axial}), geometry (i) and the different
thickness of metal and different misorientation, $T=0.05T_{C}$ and
the different thicknesses of layer normal metal between
superconductors. Spin currents $(j_{sy})_z, (j_{sz})_z$ for
geometry (i) and $(j_{sx})_z, (j_{sy})_z, (j_{sz})_z$ for geometry
(ii) are absent because of Eq.(\ref{spin}). Currents are
calculated in units of $j_{0,s}=\frac{\protect\pi }{8}\hbar
N(0)v_{F}\Delta(0)$.} \label{fig4}
\end{figure}

 In addition,  solutions of Eqs.
(\ref{Eilenberger}) satisfy the continuity conditions at the
interfaces between metal and superconductors ($z=0,l$) for all
quasiparticle
trajectories.\\
Here, as in the paper \cite{Faraii}, a simple step-like
non-self-consistent model of the constant order parameter up to
the interfaces is considered:
\begin{equation}
\mathbf{d}(z,{\mathbf{\hat{v}}}_{F})= \left\{ \begin{array}{ll}
        \mathbf{d}_1({\mathbf{\hat{v}}}_{F}) e^{+i\phi/2} & z<0 \\
        0 & 0 <z<l \\
        \mathbf{d}_2({\mathbf{\hat{v}}}_{F}) e^{-i\phi/2} & z>l
        \end{array}
        \right. ,
\end{equation}
where $\phi $ is the external phase difference between the gap
functions of superconducting bulks. We assume that the order
parameter does not depend on coordinates and in each
superconductors equals to its value far from the interface in the
left or right bulks. For such a model, the current-phase
dependence of a Josephson junction can be calculated for specific
model of $f-$wave pairing symmetry. We believe that
under this assumptions our results describe the real situation
qualitatively\cite{Coury,Likharev}. In the framework of such
model, the analytical expressions for the charge and spin currents
and density of states can be obtained for an arbitrary form of the
gap vector.

\section{Analytical Results of Green's functions}
\label{3}The solution of Eilenberger equations allows us to
calculate the charge and spin current densities in normal metal.
The expression for the charge current is:
\begin{equation}
\mathbf{j}_{e}\left( \mathbf{r}\right) =2i\pi eTN\left( 0\right)
\sum_{m}\left\langle \mathbf{v}_{F}g_{1}\left( \mathbf{\hat{v}}_{F},\mathbf{r%
},\varepsilon _{m}\right) \right\rangle,  \label{charge-current}
\end{equation}
that for the spin current is:
\begin{equation}
\mathbf{j}_{s_{i}}\left( \mathbf{r}\right) =i\pi \hbar TN\left(
0\right) \sum_{m}\left\langle \mathbf{v}_{F}\left( \mathbf{{\hat{e}}}_{i}%
\mathbf{g}_{1}\left( \mathbf{\hat{v}}_{F},\mathbf{r},\varepsilon
_{m}\right) \right) \right\rangle  \label{spin-current}
\end{equation}
and that for local density of states at the energy E is:
\begin{equation}\label{dosformula}
 \hspace{-2cm} N(E,\mathbf{r})=N\left(
0\right)\left\langle Reg_{1}\left(
\mathbf{\hat{v}}_{F},\mathbf{r},\varepsilon_m\rightarrow
  iE+\delta \right)
\right\rangle
\end{equation}
where $\left\langle ...\right\rangle $ stands for averaging over
the directions of electron momentum on the Fermi surface
${\mathbf{\hat{v}}}_{F}$, $N\left( 0\right) $ is the electron
density of states at the Fermi surface and $\mathbf{{\hat{e}}}_{i}\mathbf{=}\left( \hat{\mathbf{x}},\hat{%
\mathbf{y}},\hat{\mathbf{z}}\right) $.

\begin{figure}[tbp]
\includegraphics[width=1.1\columnwidth]{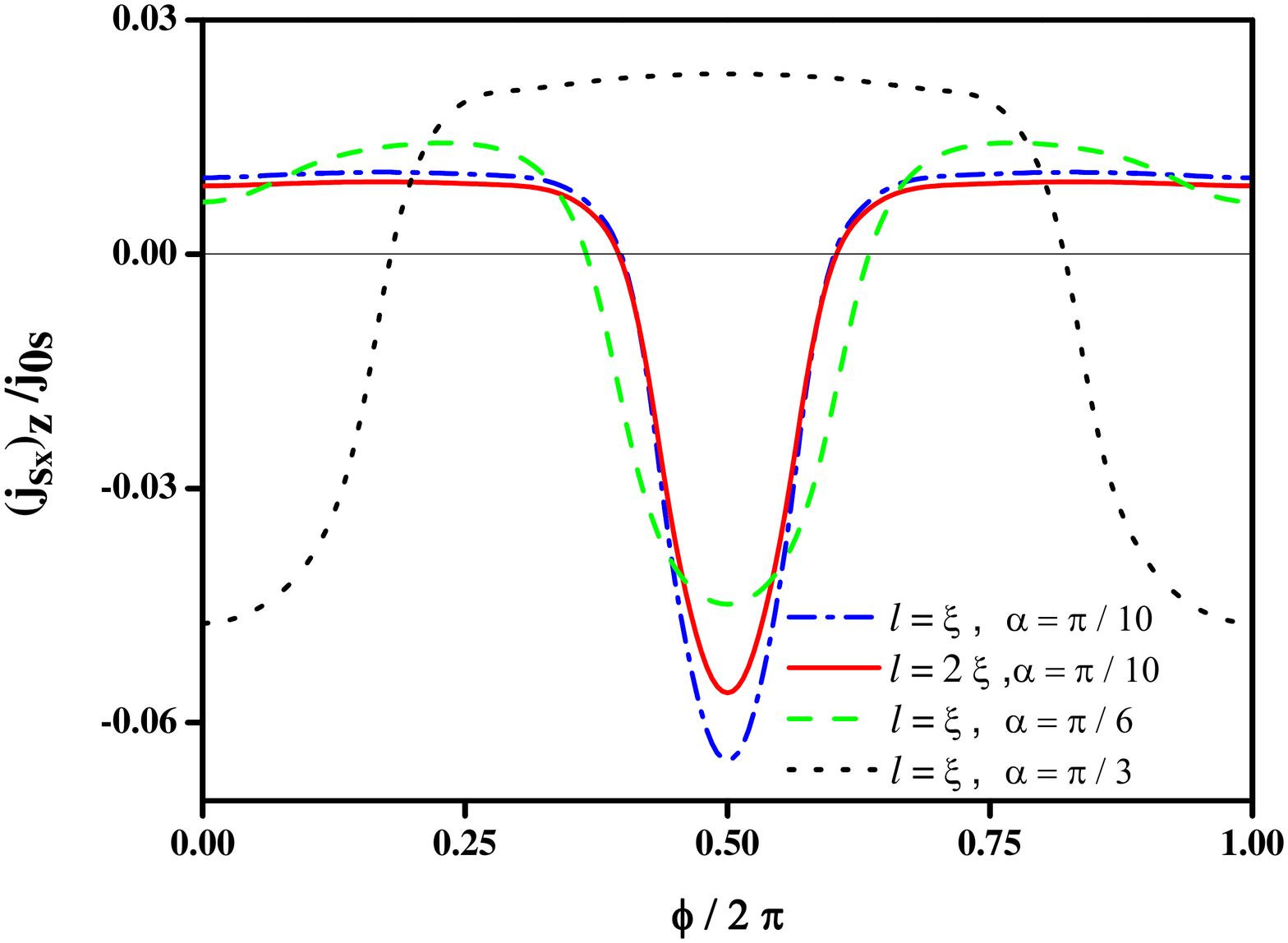}
\caption{(Color online) The normal component of the spin current
$(j_{sx})_z$ versus the phase difference $ \protect\phi $ for the
planar state (\ref{planar}), geometry (i), $T=0.05T_{C}$,
different thicknesses of layer normal metal between
superconductors and different misorientations between
superconductors. Note that spin currents
$(j_{sy})_z$ and $(j_{sz})_z$ vanish.}
 \label{fig5}
\end{figure}
\begin{figure}[tbp]
\includegraphics[width=1.1\columnwidth]{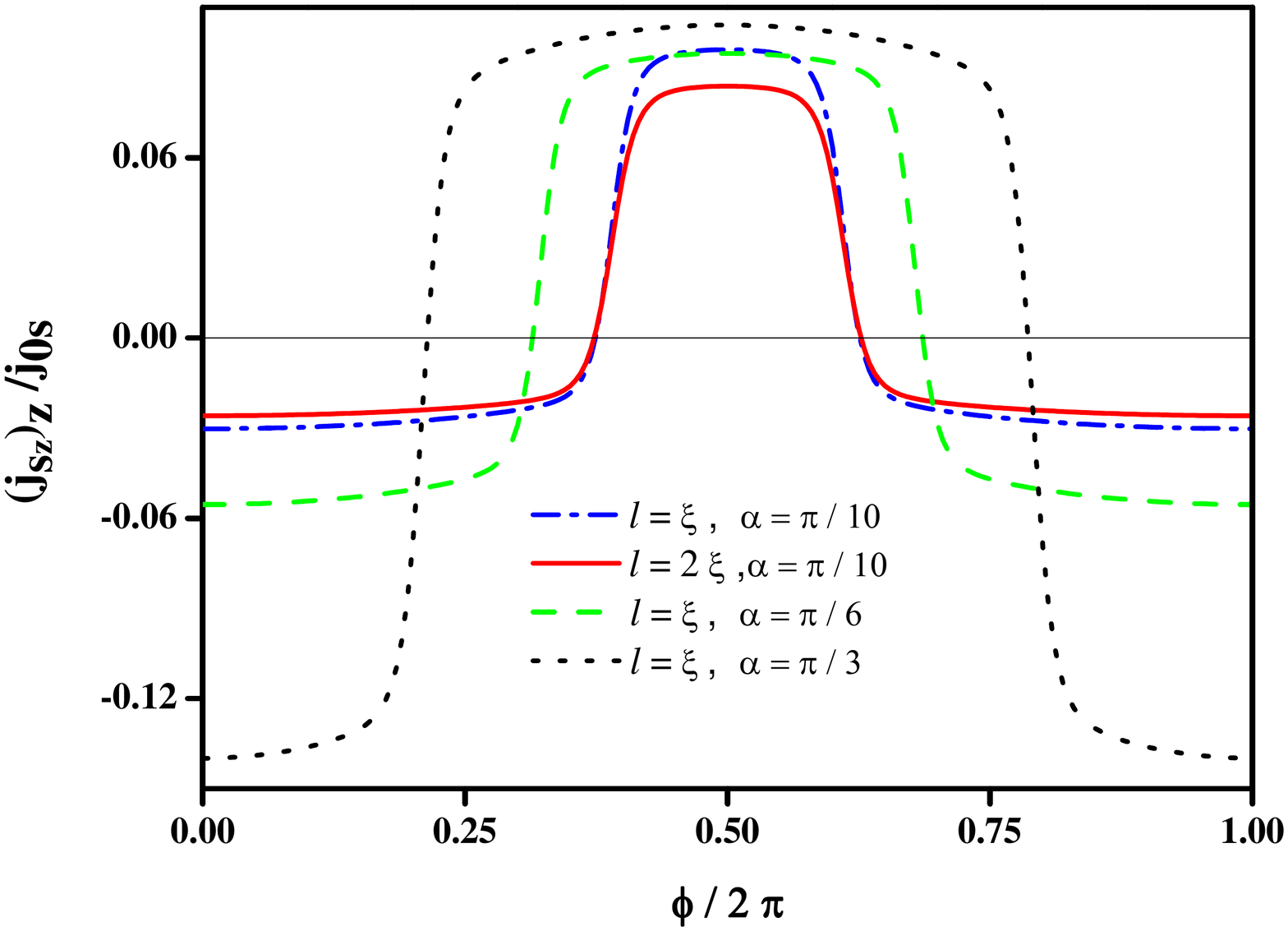}
\caption{(Color online) The normal component of the spin current
$(j_{sz})_z$ versus the phase difference $ \protect\phi $ for the
planar state (\ref{planar}), geometry (ii), $T=0.05T_{C}$,
different thicknesses of layer normal metal between
superconductors and different misorientations between
superconductors. Spin currents of $(j_{sx})_z$ and $ (j_{sy})_z$ are absent.
}
 \label{fig6}
\end{figure}
\begin{figure}[tbp]
\includegraphics[width=\columnwidth]{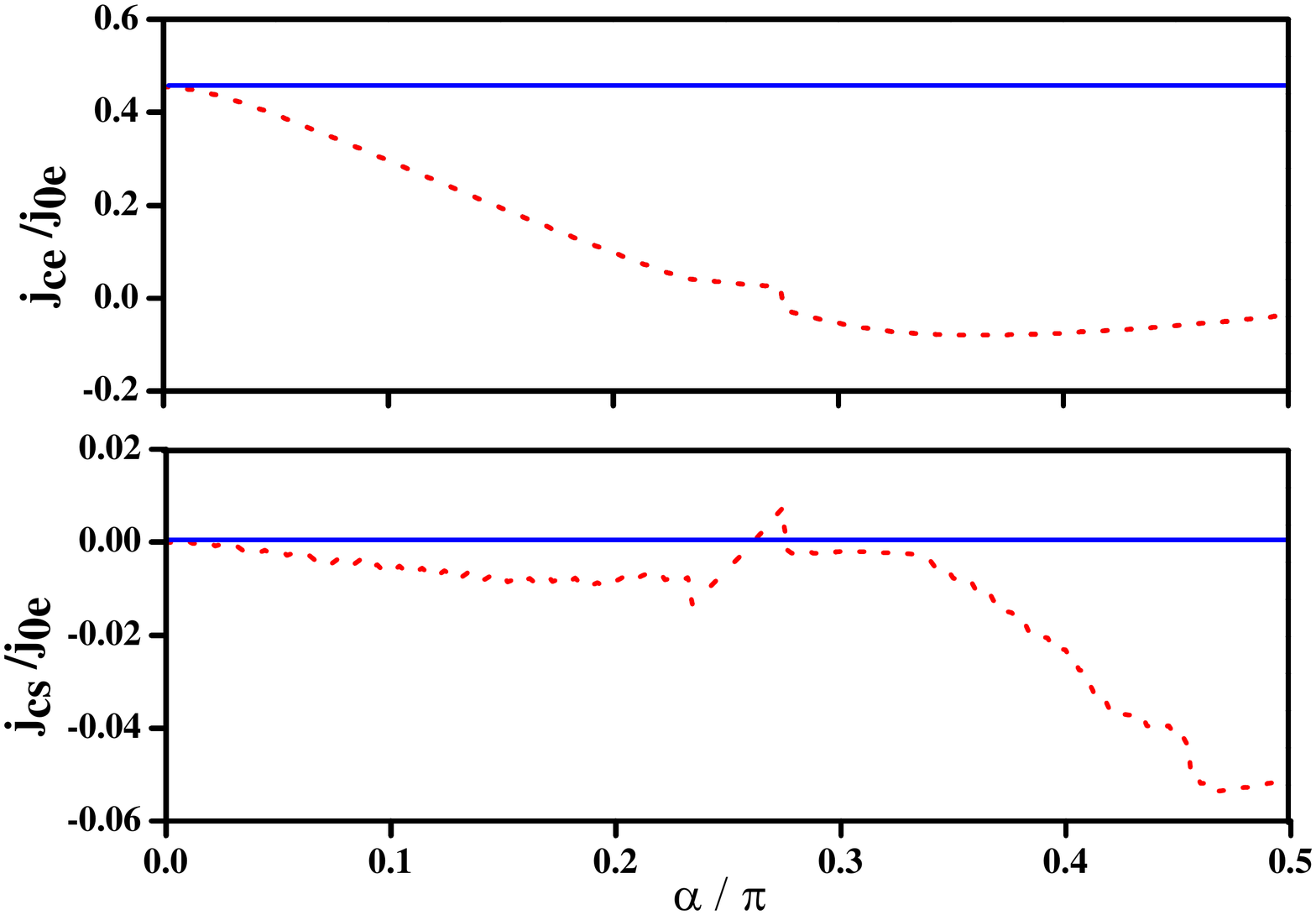}
\caption{(Color online) Critical currents of charge and spin
$(j_{sx})_z$ as a function of misorientation angle for axial state
(\ref{axial}), $l/\xi=0.5$, $T=0.05T_c$. Solid lines are for
geometry (ii) and dotted lines are for geometry (i). Here and in
Fig.\ref{fig8}, we define
$j_{ce}=Max_{\phi}j_{e}(\phi)=j_{e}(\phi^{\ast})$ and
$j_{cs}=j_{s}(\phi^{\ast})$.} \label{fig7}
\end{figure}
\begin{figure}[tbp]
\includegraphics[width=\columnwidth]{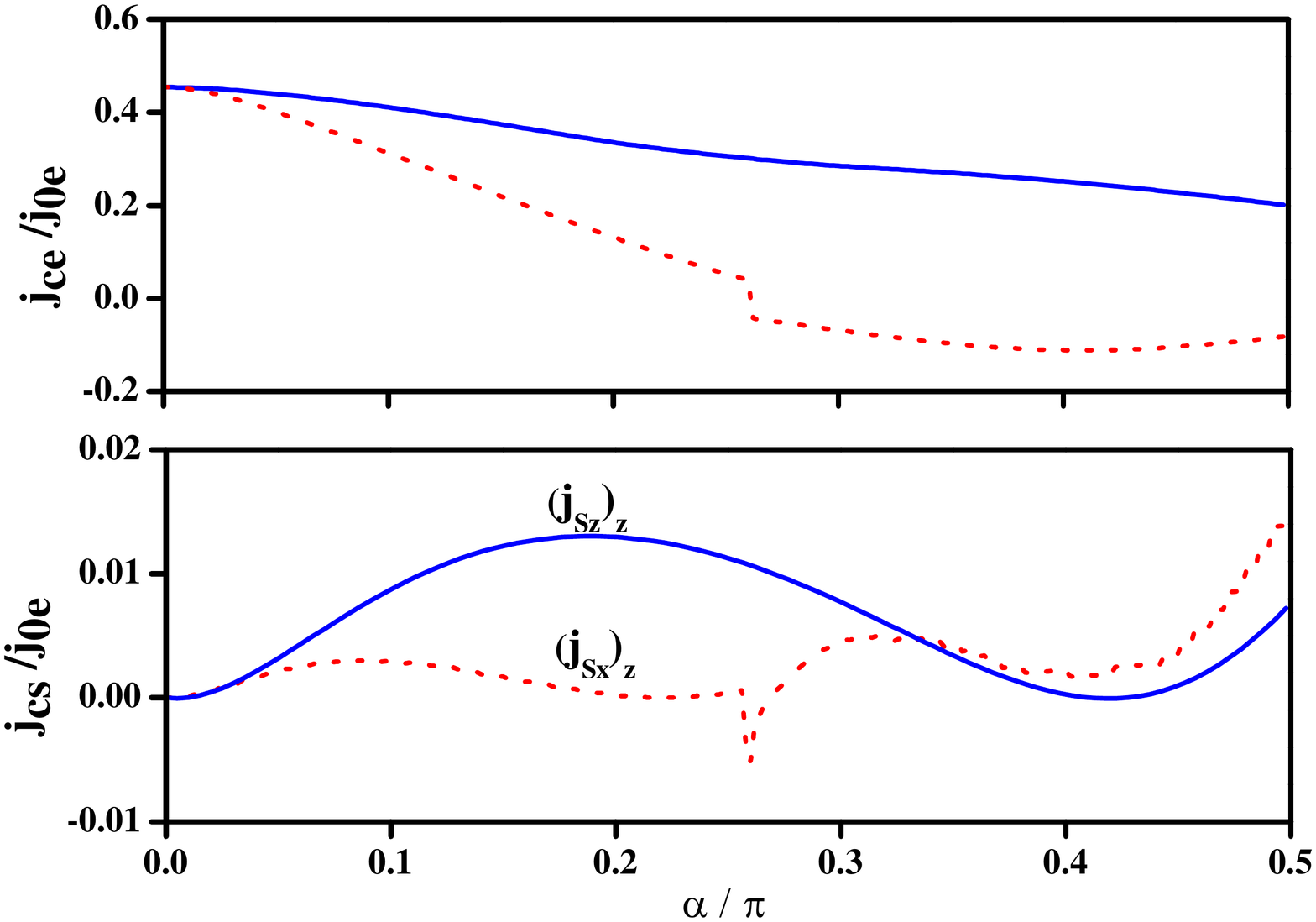}
\caption{(Color online) Critical charge and spin currents as a
function of misorientation angle for planar state (\ref{planar}),
$l/\xi=0.5$, $T=0.05T_c$. Solid lines are for geometry (ii) and
dotted lines are for geometry (i). } \label{fig8}
\end{figure}

\begin{figure}[tbp]
\includegraphics[width=1.1\columnwidth]{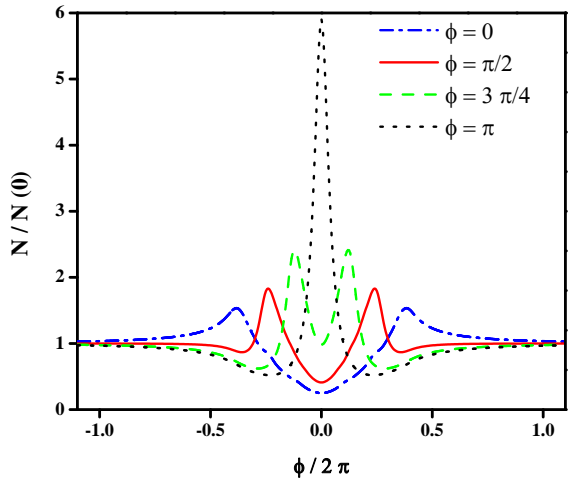}
\caption{(Color online) Local density of states in the normal
metal for $\alpha=0$, $l/\xi=0.5$, $T=0.05T_c$ and for both
symmetries (axial and planar states) and different phases $\phi$.
} \label{fig9}
\end{figure}

\begin{figure}[tbp]
\includegraphics[width=1.1\columnwidth]{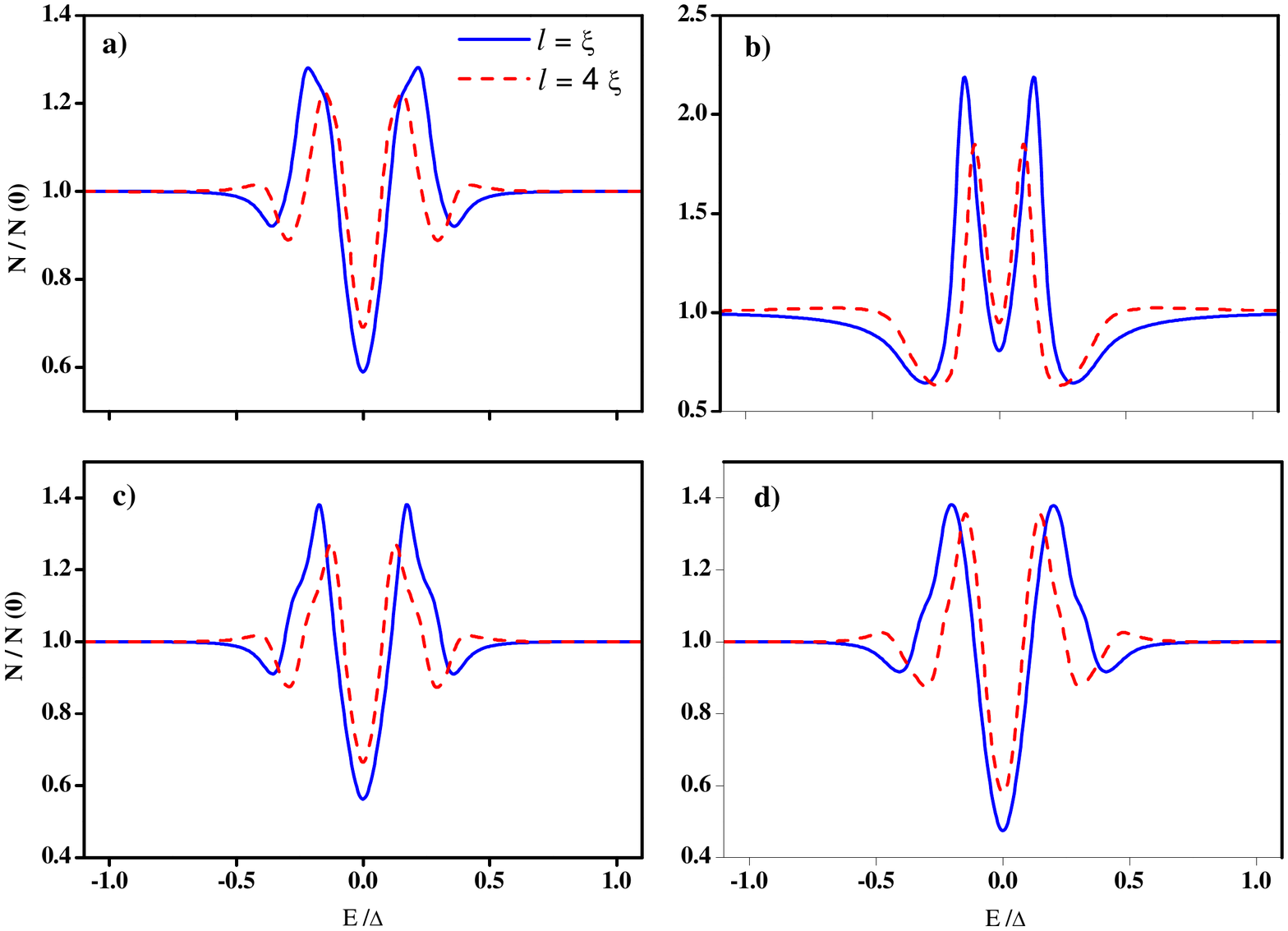}
\caption{(Color online) Local density of states in the normal
metal for both geometries with $\alpha=\frac{\pi}{10}$,
$\phi=\frac{\pi}{2}$, $T=0.05T_c$, and different thicknesses of
normal metal $l$. Panels (a) and (b) are for axial state
(\ref{axial}) and panels (c) and (d) are for planar state
(\ref{planar}).
 Left plots are for geometry (i) and right plots are for geometry (ii). } \label{fig10}
\end{figure}

\begin{figure}[tbp]
\includegraphics[width=1.1\columnwidth]{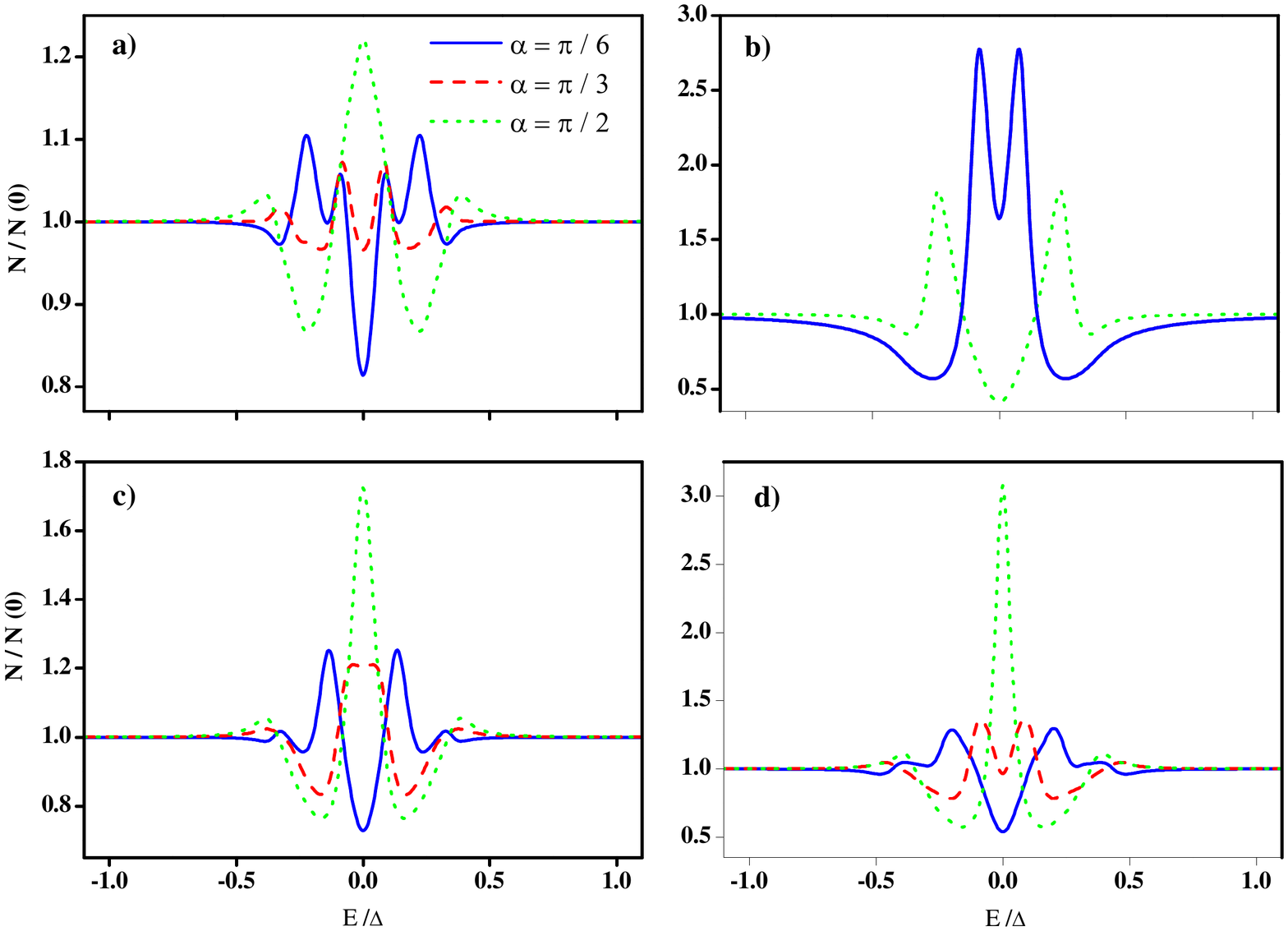}
\caption{(Color online) Local density of state in the normal metal
for $l/\xi=0.5$, $T=0.05T_c$, $\phi=\frac{\pi}{2}$, and two
different misorientations. Panels (a) and (b) are for axial state
(\ref{axial}) and panels (c) and (d) for planar state
(\ref{planar}).
 Left plots are for geometry (i) and right plots are for geometry (ii).
 In panel (b), plots of $\alpha=\frac{\pi}{3}$ and $\alpha=\frac{\pi}{6}$ (which are coincident) are shown by
 a solid line.} \label{fig11}
\end{figure}

The calculated Green's functions in the normal metal
are the following:
\begin{equation}\label{charge}
g_{1N}=\frac {\eta( A-B )} {A+B+2|\mathbf{d_1}\cdot
\mathbf{d_2^*}|^{2}}
\end{equation}
\begin{equation}
\mathbf{g}_{1N}=\frac {2\eta\mathbf{d_1^*}\cdot\mathbf{d_2}
(\mathbf{d_1}\times\mathbf{d^*_2}) } {A+B+2|\mathbf{d_1}\cdot
\mathbf{d_2^*}|^{2}}\label{spin}
\end{equation}
where $\eta =sgn\left( v_{z}\right)$,
$\Omega_{n}=\sqrt{\varepsilon_{m}^{2}+\left|\mathbf{d}_{n}\right|^{2}}$,
\begin{equation}\label{def green}
A=\mathbf{d_1^*}\cdot\mathbf{d_2}(\varepsilon_m+\eta\Omega_1)(\varepsilon_m+\eta\Omega_2)\exp(\frac{+2\varepsilon_m
l}{|v_z|} )
\end{equation}
and
\begin{equation}
B=\mathbf{d_1}\cdot\mathbf{d_2^{*}}(\varepsilon_m-\eta\Omega_1)(\varepsilon_m-\eta\Omega_2)\exp(\frac{-2\varepsilon_m
l}{|v_z|} ).
\end{equation}
Using these Green's function, we obtain the charge and
spin current through the SNS Josephson junction and also the density of
states of system in the normal region $0\leq z\leq l$.

\section{Numerical Results of Currents and Density of States}
\label{4} In this section we calculate charge and spin currents in SNS
junctions having in mind UPt$_3$ that is a $f-$wave superconductor. Two
famous models for order parameter of UPt$_3$ are considered.
The first is the axial state \cite{Matthias}:
\begin{equation}
\mathbf{d}(T,\mathbf{v}_{F})=\Delta(T)\hat{\mathbf{z}}\
k_{z}(k_{x}+ik_{y})^2, \label{axial}
\end{equation}
and the other is the planar state \cite{Machida}:
\begin{equation}
\mathbf{d}(T,\mathbf{v}_{F})=\Delta(T)k_{z}[\hat{\mathbf{x}}\
(k_{x}^2-k_{y}^2)+\hat{\mathbf{y}}2k_xk_y]. \label{planar}
\end{equation}
The coordinate axes $\hat{%
\mathbf{x}},\hat{\mathbf{y}},\hat{\mathbf{z}}$ are chosen along
the crystallographic axes
$\hat{\mathbf{a}},\hat{\mathbf{b}},\hat{\mathbf{c}}$ in the left
side of Fig.\ref{fig1}. The function $\Delta\left( T\right) $ in Eqs.(\ref{axial}) and (\ref{planar})
describes the dependence of the order parameter $\mathbf{d}$ on
the temperature $T$. Using Green's function in
Eqs.(\ref{charge}) and (\ref{spin}), we have calculated charge and
spin currents, and the density of states in $f-$wave Josephson junction with pairing symmetry of
Eqs. (\ref{axial}) and (\ref{planar}) numerically.\\

Two superconducting bulks may have misorientation which has been
created by geometries (i) and (ii) in Fig. \ref{fig1}. For two
geometries and specific models of $f-$wave pairing symmetry we
have plotted the currents in terms of phase difference in
Figs.\ref{fig2}-\ref{fig6}. The obtained currents are periodic
functions of phase difference between superconductors with a
period of $2\pi$. It is clarified that by increasing the thickness
of normal layer, amplitude of currents decreases, which is
understandable because by increasing thickness of the normal metal,
quantum coherency of macroscopic phases between left and right
superconductors decreases. The current-phase relations are
qualitatively different from the case of $s-$wave \cite{Kulik} and
$d-$wave Josephson junction \cite{Barash,Coury}. In particular, an
interesting case is the zero of charge current at finite phase
$\phi_0$ in Fig.\ref{fig2} (b) which occurs for the most of
junctions exactly at $\phi=0$ or $\phi=\pi$. On the other hand,
for finite misorientation between left and right gap vectors even
at the zero phase difference we obtain finite charge current. This
is because according to Eq. (\ref{axial}) in geometry (ii)
misorientation $\alpha$ plays role of phase difference --rotation
by $\alpha$ gives the phase factor $\exp(2i\alpha)$. Also, in
Figs. \ref{fig4}-\ref{fig6}, it is found that the spin current of
Josephson junction between two triplet superconductors has been
created by misorientation between gap vectors of superconductors
\cite{Kolesnichenko}. For the case of geometry (ii) and axial
state (\ref{axial}), spin current is absent because both of gap
vectors are in the same direction. Also, by using Eq.
(\ref{spin}), we find that spin currents $(j_{sy})_z$ and
$(j_{sz})_z$  for the axial state (12) and also the planar state
(13) in geometry (i), and spin currents  $(j_{sx})_z$ and $
(j_{sy})_z$ for the planar state (13) in geometry (ii) are absent.
In contrast to the charge current, the spin current has a large
value at $\phi=\pi$ (or at $\phi=0$ for large misorientation
angle). 
It is seen that charge current are generally odd function but spin current are
even function of the phase difference because charge current is
odd in time reversal while spin current is even. 
Since spin current is an even function of phase, its derivative becomes odd in phase and hence should be zero at $\phi=0$ and $\pi$. Thus, the spin current has a local maximum or minimum at $\phi=0$ and $\pi$.
We also see that for some phases as $\phi=\pi$ the spin
current exists but charge current disappears. This is a purely
spin transport in the absence of charge current. 
This effect can
not be observed in singlet superconducting junction like
conventional or high $T_c$ superconductors. Only in the case of
superconductor-ferromagnet-superconductor junction with
inhomogeneous magnetization this spin current is present
\cite{Linder}.

The critical currents as a function of the misorientation angle
are shown in Figs.\ref{fig7} and \ref{fig8}. Critical charge and spin
 currents, defined as
$j_{ce}=Max_{\phi}j_{e}(\phi)=j_{e}(\phi^{\ast})$ and
$j_{cs}=j_{s}(\phi^{\ast})$, respectively, are plotted as a function of
misorientation angle $\alpha$ for specified temperature and normal layer
thickness
 ($T=\frac{T_c}{20}$, $l=\frac{\xi}{2}$).
 By changing the misorientation, we see the $0-\pi$ transition due to the anisotropy of the gap.\cite{Barash}
Also, at the $0-\pi$ transition point for both symmetries in geometry
(i) we have a jump of spin current due to the jump of the phase as found in Ref.\cite{Linder} but in geometry (ii) there aren't any $0-\pi$ transition and hence jump of spin current.

Finally, in Figs.\ref{fig9}-\ref{fig11} we have plotted the
density of states in the normal region. It is known that
proximity effect in the SNS junction strongly changes the density
of states in the normal metal. In our calculations, we obtained
Andreev bound states in the junctions as shown in Figs.\ref{fig9}-\ref{fig11}. This is understandable, by
interference effect of electrons and holes in the normal metal. An
electron will be reflected by Andreev reflection as a hole. This
process is also repeated on the other SN interface. Then,
interference of holes and electrons increases the effective
Andreev reflection probability and proximity effect at the low
energies.

In Fig.\ref{fig9}, we obtain a mini-gap like
$s-$wave and $d-$wave superconductors \cite{Shumeiko}.
Exactly at $\phi=\pi$ we obtain a peak at the Fermi energy. Here,
like Josephson junction between other types of unconventional
superconductors \cite{Shumeiko}, we obtain the finite density of
states for energies inside the gap ($-\Delta <E<\Delta$).
The results of a finite misorientation and different $l$ are shown
in Fig. \ref{fig10}. The density of states gets smeared by
increasing $l$. In Fig. \ref{fig11} we plotted the density of
states for both of geometries and both of symmetries. It is
observed that by increasing the misorientation two peaks are
gradually merged into one and there will be a peak at zero energy
at $\alpha=\frac{\pi}{2}$. 
The increase of the density of states at low energy corresponds to that of the spin current. 
We find that effect of thickness of normal layer is only
quantitative but misorientation changes currents and the density
of states qualitatively.

\section{Conclusions}\label{5}
In this paper, we have investigated transport properties and also local
density of states in triplet SNS Josephson junction with two $f-$wave
superconductors. A spin-polarized current normal to the interface is
investigated theoretically. It is observed that
normal layer decreases both of charge and spin currents. It is also
found that misorientation of $d$-vector between left and right
superconducting bulks produces the spin-polarized current which
can flow even in the absence of charge current. We also unveiled a jump of the spin current associated with the 0-$\pi$ transition. 
In addition, density of states have been calculated for various
phase difference, normal metal thickness and misorientation.
Normal metal thickness has only quantitative effect on the mid-gap
states. Misorientation of the $d$-vectors drastically changes the
junction property.

A generalization of calculations  of this paper to
superconductor-ferromagnet-superconductor junctions is an
interesting subject. We will do this generalization in the next
work.


\begin{thebibliography}{99}
\bibitem{Maeno} Y. Maeno, H. Hashimoto, K. Yoshida, S. Nishizaki, T. Fujita, J. G. Bednorz, and F. Lichtenberg, Nature (London) \textbf{372}, 532 (1994).

\bibitem{Ishida} K. Ishida, H. Mukuda, Y. Kitaoka, K. Asayama, Z. Q. Mao, Y. Mori, and Y. Maeno, Nature (London) \textbf{396}, 658 (1998).

\bibitem{Luke} G. M. Luke, Y. Fudamoto, K. M. Kojima, M. I. Larkin, J. Merrin, B. Nachumi, Y. J. Uemura, Y. Maeno, Z. Q. Mao, Y. Mori, H. Nakamura, and M. Sigrist, Nature (London) \textbf{394}, 558 (1998).


\bibitem{Mackenzie} A. P. Mackenzie and Y. Maeno, Rev. Mod. Phys. \textbf{75}, 657 (2003).
\bibitem{Nelson} K. D. Nelson, Z. Q. Mao, Y. Maeno and Y. Liu, Science \textbf{306}, 1151 (2004).
\bibitem{Asano1} Y. Asano, Y. Tanaka, M. Sigrist and S. Kashiwaya, Phys. Rev. B \textbf{67}, 184505 (2003);
 Phys. Rev. B \textbf{71}, 214501 (2005).
\bibitem{Tou} H. Tou, Y. Kitaoka, K. Ishida, K. Asayama, N. Kimura, Y. Onuki, E. Yamamoto, Y. Haga and
K. Maezawa, Phys. Rev. Lett. \textbf{80}, 3129 (1998).
\bibitem{Muller} V. Muller, Ch. Roth, D. Maurer, E. W. Scheidt, K. Lers, E. Bucher and H. E. Bmel, Phys. Rev. Lett. \textbf{58}, 1224 (1987).
\bibitem{Qian} Y. J. Qian, M. F. Xu, A. Schenstrom, H. P. Baum, J. B. Ketterson, D. Hinks,
M. Levy and B. K. Sarma, Solid State Commun., \textbf{63}, 599
(1987).
\bibitem{Abrikosov} A. A. Abrikosov, J. of Low Temp. Phys.  \textbf{53}, 359 (1983).
\bibitem{Fukuyama} H. Fukuyama and Y. Hasegawa, J. Phys. Soc. Jpn \textbf{56}, 877 (1987).
\bibitem{Lebed} A. G. Lebed, K. Machida and M. Ozaki, Phys. Rev. B \textbf{62}, 795 (2000).

\bibitem{Saxena} S. S. Saxena, P. Agarwal, K. Ahilan, F. M. Grosche, R. K. W. Haselwimmer, M. J. Steiner, E. Pugh, I. R. Walker, S. R. Julian, P. Monthoux, G. G. Lonzarich, A. Huxley, I. Shelkin, D. Braithwaite, and J. Flouquet, Nature (London) \textbf{406}, 587 (2000).

\bibitem{Pfleiderer} C. Pfleiderer, M. Uhlarz, S. M. Hayden, R. Vollmer, H. v. Lohneysen, N. R. Bernhoeft, and G. G. Lonzarich, Nature (London) \textbf{412}, 58 (2001).

\bibitem{Aoki} D. Aoki, A. Huxley, E. Ressouche, D. Braithwaite, J. Flouquet, J. Brison, E. Lhotel, and C. Paulsen, Nature (London) \textbf{413}, 613 (2001).

\bibitem{Graf} M. J. Graf, S. K. Yip and J. A. Sauls, Phys. Rev. B, \textbf{62}, 14393 (2000).
\bibitem{Machida} K. Machida, T. Nishira, and T. Ohmi, J. Phys. Soc.
Jpn, \textbf{68}, 3364 (1999).
\bibitem{Lussier} B. Lussier,  B. Ellman and L. Taillefer , Phys. Rev. B, \textbf{53}, 5145 (1996).
\bibitem{Heffner} R. H. Heffner and M. R. Norman, Comment on Cond. Matt. Phys. \textbf{17}, 361 (1996).

\bibitem{Tanaka} Y. Tanaka, Y. V. Nazarov, and S. Kashiwaya, Phys. Rev. Lett. \textbf{90}, 167003 (2003);  Y. Tanaka, Yu. V. Nazarov, A. A. Golubov, and S. Kashiwaya, Phys. Rev. B \textbf{69}, 144519 (2004).
\bibitem{Tanaka2} Y. Tanaka, S. Kashiwaya, and T. Yokoyama, Phys. Rev. B \textbf{71}, 094513 (2005).
\bibitem{Tanuma} Y. Tanuma, Y. Tanaka, and S. Kashiwaya, Phys. Rev. B \textbf{74}, 024506 (2006).

\bibitem{Asano2} Y. Asano, Y. Tanaka, and S. Kashiwaya, Phys. Rev. Lett. \textbf{96}, 097007 (2006); Y. Asano, Y. Tanaka, T. Yokoyama, and S. Kashiwaya, Phys. Rev. B \textbf{74}, 064507 (2006).

\bibitem{Yokoyama2} T. Yokoyama, Y. Tanaka, A. A. Golubov, and Y. Asano, Phys. Rev. B \textbf{73}, 140504(R) (2006); T. Yokoyama, Y. Tanaka, and A. A. Golubov, Phys. Rev. B \textbf{75}, 094514 (2007).

\bibitem{Sawa} Y. Sawa, T. Yokoyama, Y. Tanaka, and A. A. Golubov, Phys. Rev. B \textbf{75}, 134508 (2007).


\bibitem{Guttman} G. D. Guttman, E. B. Jacob and D. J. Bergman, Phys. Rev. B \textbf{57}, 2717 (1998).
\bibitem{Lindell} R. Lindell, J. Penttila, M. Sillanpaa and P. Hakonen , Phys. Rev. B \textbf{68}, 052506 (2003).
\bibitem{Golubov} A. A. Golubov, M. Yu. Kupriyanov and E. Ilichev, Rev. Mod. Phys \textbf{76}, 411 (2004).
\bibitem{Kolesnichenko} G. Rashedi and Yu. A. Kolesnichenko, Physica c, \textbf{451}, 31 (2007).
\bibitem{Kulik} I. O. Kulik and A. N. Omelyanchouk, Fiz. Nizk. Temp \textbf{4}, 296 (1978)
[Sov. J. Low Temp. Phys., \textbf{4}, 142 (1978)].
\bibitem{Rashedi} G. Rashedi and Yu. A. Kolesnichenko, Phys. Rev. B \textbf{69}, 024516 (2004).

\bibitem{Asano3} Y. Asano, Phys. Rev. B \textbf{72}, 092508 (2005); Phys. Rev. B \textbf{74}, 220501(R) (2006).

\bibitem{Barash} Yu. S. Barash, H. Burkhardt, and D. Rainer, Phys. Rev. Lett. \textbf{77}, 4070 (1996); Y. Tanaka and S. Kashiwaya, Phys. Rev. B \textbf{53}, R11957 (1996); \textit{ibid}. \textbf{56}, 892 (1997).

\bibitem{Coury} M. H. S. Amin, M. Coury, S. N. Rashkeev, A. N. Omelyanchouk and
A. M. Zagoskin, Physica B \textbf{318}, 162 (2002).
\bibitem{Likharev} K. K. Likharev, Rev. Mod. Phys \textbf{51}, 101 (1979).
\bibitem{Eilenberger} G. Eilenberger, Z. Phys, \textbf{214}, 195 (1968).
\bibitem{Faraii} Z. Faraii and M. Zareyan, Phys. Rev. B \textbf{69}, 014508 (2004).
\bibitem{Matthias} J. G. Matthias, S. K. Yip and J. A. Sauls, Phys.
Rev. B \textbf{62}, 14393 (2000).
\bibitem{Linder} M. Alidoust, J. Linder, G. Rashedi, T. Yokoyama and Asle
Sudbo, Phys. Rev. B \textbf{81}, 014512 (2010).
\bibitem{Shumeiko} T. Lofwander, V. S. Shumeiko and G. Wendin, Supercond. Sci. Technol. \textbf{14}, R53
(2001).
\end{thebibliography}
\end{document}